\documentclass[10pt]{iopart}
\usepackage{graphicx}
\usepackage{latexsym}
\usepackage{dcolumn}
\usepackage{bm}
\usepackage{color}
\usepackage{amssymb}
\usepackage{amsmath}

\newcommand{\be}{\begin{equation}}
\newcommand{\ee}{\end{equation}}
\newcommand{\bea}{\begin{eqnarray}}
\newcommand{\eea}{\end{eqnarray}}
\newcommand{\bse}{\begin{subequations}}
\newcommand{\ese}{\end{subequations}}
\newcommand{\p}{\partial}

\renewcommand{\e}{\epsilon}

\newcommand{\commento}[1]{}
\newcommand{\oz}{\omega_0}

\begin{document}

\paper[]{Frozen dynamics of a breather induced by an adiabatic invariant}
\date{\today}

\author{Antonio Politi$^{1}$, Paolo Politi$^{2,3}$, Stefano Iubini$^{2,3}$}

\address{$^{1}$ Institute for Complex Systems and Mathematical Biology \& SUPA
           University of Aberdeen, Aberdeen AB24 3UE, United Kingdom}
\address{$^{2}$ Istituto dei Sistemi Complessi, Consiglio Nazionale
delle Ricerche, via Madonna del Piano 10, I-50019 Sesto Fiorentino, Italy}
\address{$^{3}$ Istituto Nazionale di Fisica Nucleare, Sezione di Firenze, via G. Sansone 1 I-50019, Sesto Fiorentino, Italy}
\ead{a.politi@abdn.ac.uk, paolo.politi@cnr.it, stefano.iubini@cnr.it}

\begin{abstract}
The Discrete Nonlinear Schr\"odinger (DNLS) equation is a Hamiltonian model
displaying an extremely slow relaxation process when discrete breathers appear in the system.
In [Iubini S, Chirondojan L, Oppo G L, Politi A and Politi P 2019 
\textit{Physical Review Letters} \textbf{122} 084102], it was conjectured that the 
frozen dynamics of tall breathers is due to the existence of an adiabatic invariant (AI).
Here, we prove the conjecture in the simplified context of a unidirectional 
DNLS equation, where the breather is ``forced" by a background unaffected by
the breather itself.
We first clarify that the nonlinearity of the breather dynamics and the deterministic nature of
the forcing term are both necessary ingredients for the existence of a frozen dynamics.
We then derive perturbative expressions of the AI by implementing a
canonical perturbation theory and via a more phenomenological approach based on the
estimate of the energy flux.
The resulting accurate identification of the AI allows revealing the presence and role of sudden jumps as 
the main breather destabilization mechanism, with an
unexpected similarity with L\'evy processes.
\end{abstract}
\noindent{\bf Keywords:} Hamiltonian systems; Frozen dynamics; Breathers; Adiabatic invariant

\submitto{Journal of Statistical Mechanics: theory and experiment}

\maketitle

\section{Introduction}
\label{sec:intro}

The dynamical slowing down of relaxation in a macroscopic system has received great attention in the literature
since the pioneering numerical experiments of Fermi, Pasta, Ulam and Tsingou (FPUT)~\cite{fermi1955}. 
In this case, the low-temperature slow dynamics and the related ergodicity-breaking mechanism can be 
attributed to the vicinity of the FPUT model to the integrable Toda chain~\cite{BCP13}.
In the opposite high-temperature limit, energy concentration on a single or few lattice sites has been found
to produce similar slowing down effects. These are due to effective decoupling between the
localized hot spot and the rest of the system. 
As an example, in the rotor chain,
it was found that in this limit, ergodization time becomes anomalously larger than 
the Lyapunov time, i.e. the characteristic chaotic time scale~\cite{mithun19}.
 Similar nonergodic behavior was also found in the Klein-Gordon lattice~\cite{danieli19} and in 
 a harmonic model with hard-walls pinning potential~\cite{hahn21}.

In this paper, we focus on the Discrete Nonlinear Schr\"odinger (DNLS) equation~\cite{kevrekidis09}, a classical Hamiltonian model
used to study nonlinear propagation in discrete media when dissipation is negligible.
This includes electronic transport in biomolecules, wave propagation in photonic crystals, trapped ultracold gases, and magnetic systems.
The presence of two conservation laws (mass density $a$ and energy density $h$, see Sec.~\ref{sec.uni} for a precise definition) 
determines a non trivial microcanonical phase diagram, characterized by an infinite temperature-line at finite energy, $h_c(a)=2a^2$. 
The first study of the equilibrium properties of the DNLS equation is due to Rasmussen et al.~\cite{RCKG00},
who discuss the statistical mechanics for $h\leq h_c$ within the grandcanonical ensemble
(more recent studies can be found in Refs.~\cite{levy18,levy21a,levy21b,chatterjee21}). 

Below $h_c$, standard equipartition and equivalence of ensembles occur and equilibrium states are characterized by a positive temperature 
$0\leq T <\infty$. Above $h_c$, equipartition is broken by the emergence of spatially localized discrete breathers on top of a chaotic background. 
In this region, 
the system is expected to relax towards an equilibrium state where a finite fraction
of the whole energy, equal to $(h-h_c)/h$, is localized on a single site~\cite{R1,R2,R3,R4}.
Hence, it is natural to interpret $h_c(a)$ as a condensation transition-line.
Numerical microcanonical simulations, however, have revealed the existence of a region above $h_c$,
characterized by a negative microcanonical temperature, where 
breathers are continuously born and die~\cite{IFLOP}.
The very existence of this seemingly stationary behavior, although confirmed in Ref.~\cite{mithun18}, 
is challenged by the theoretical analysis performed in the limit of a negligible interaction
energy~\cite{GILM1,GILM2}. In fact, by means of equilibrium large-deviations techniques, it has been proven that stationary
delocalized states persist up to $h^*=2a^2 + c_1/N^{1/3}$ ($c_1$ being a positive constant).
Accordingly, stationary ergodic states are a finite-size effect nevertheless observable for large system
sizes and associated to negative microcanonical temperatures.

If the equilibrium properties of the DNLS equation are more or less understood,
nonequilibrium dynamics and relaxation processes are much less so.
Simplified stochastic versions of the DNLS equation show a condensation process occurring
via a coarsening dynamics~\cite{IPP14}  where the ``extra energy"
is concentrated in a certain number of localization sites, whose density $\rho$ decreases as a power-law
with a subdiffusive exponent~\cite{IPP17}.
On the other hand, numerical simulations of the full DNLS equation, performed above 
the critical line show an almost frozen state rather than
a coarsening process. 

Studying dynamics for $h>h_c$ is not easy because of the strong
finite-size corrections, which induce a delocalized phase up to $h^*$.
Since  the slowness of the relaxation processes is essentially due to the weak exchange of energy
of the breathers with the surroundings, it is more convenient to analyze the 
relaxation of a breather interacting with a thermal background in the  region $0< T< \infty$. 
Using this setup, i.e. studying the relaxation time $\tau_{b_0}$ of a single breather of initial mass $b_0$,
it was found~\cite{ICOPP} that $\tau_{b_0}$ increases exponentially with $b_0$, $\tau_{b_0} \simeq \exp(c b_0)$.
Long breather life-times in the DNLS equation are known since a long 
time~\cite{johansson01}, but the earlier studies 
focused exclusively on a setup where the background is very small and thereby characterized by
a regular dynamics. Hence the breather stability is controlled by entirely different mechanisms.

In Ref.~\cite{ICOPP}, it was conjectured that the slow relaxation follows from the presence of an 
adiabatic invariant (AI), i.e. of a quasi-conserved quantity.
The idea is natural, since we are in the presence of a Hamiltonian system with two well separated time scales:
that of the chaotic background activity and the fast rotation of the breather.
However, in Ref.~\cite{ICOPP} we were unable to substantiate our conjecture because
we limited to identify the lowest-order approximation of the AI with the breather energy.

In this paper we revisit the problem, by considering a simplified setup
where the coupling between the breather and the rest of the
system, i.e. the incoherent background at $T>0$, is unidirectional (UC model): this means that the breather feels the evolution of the
background but not vice-versa. The UC model has the same frozen dynamics as the full DNLS, therefore
reassuring about the generality of its results.
More precisely, using the UC model we are able to demonstrate that slow dynamics appearing in DNLS 
is due to the presence of an AI.
This result is obtained in two steps:
first, we clarify the role of the stochastic-like dynamics of the background, as well as the nonlinear nature of
the breather rotations;
second, we perform a rigorous perturbative analysis, deriving explicit
expressions of the AI at higher orders of approximation. 
This analysis is also possible because we can perform faster simulations, allowing a precise test
of our predictions. 

A summary of the content of the paper follows.
In Sec.~\ref{sec.uni} we  introduce the full DNLS model and its unidirectional counterpart,
and we justify the UC model.
In Sec.~\ref{sec.necessary} we show that the nonlinear as well as the deterministic character of the
model are necessary ingredients to ensure frozen dynamics.
In Sec.~\ref{sec.sufficient} we present two methods to derive perturbative expressions of the AI in the UC model.
The former one is based on the canonical perturbation theory. The latter is based on the reformulation of the
energy flux (out of the breather) as the time derivative of a stationary function.
In Sec.~\ref{sec.results} we analyze the different perturbative orders of the AI and make use of the AI expression to
analyse numerically the diffusion of the breather energy.
In the last Sec.~\ref{sec.disc} we discuss the main results and mention the open questions.

\section{From the full DNLS model to its unidirectional version}
\label{sec.uni}

The DNLS equation is a phenomenological model having the form
\begin{equation}
i \dot {z}_n = -2 |z_n|^2z_n - z_{n+1}-z_{n-1} \equiv - \frac{\p H}{\p z_n^*} \; ,
\label{eq:dnls}
\end{equation}
where $z_n$ are complex variables, $n$ is the index of the lattice site, and
\begin{equation}
 H= \sum_n \left( |z_n|^4+z_n^*z_{n+1}+z_nz_{n+1}^* \right) 
 \label {Hz}
 \end{equation}
is the total energy, which is conserved. There is a second exactly conserved quantity,
the total mass $A=\sum_n |z_n|^2$, because of the invariance of $H$ under  global
phase rotations ($z_n \to z_n e^{i\bar\phi}$).

If $h=H/N$ and $a=A/N$ ($N$ being the total number of sites) are respectively the energy and mass densities,
it is known that $h= h_0 \equiv a^2 -2a$ corresponds to zero-temperature states, while along the line
$h=h_c \equiv 2a^2$ the temperature is infinite. Finite, positive temperatures are found in between the two lines.
The relationship between mass-energy $(a,h)$ and chemical potential-temperature $(\mu,T)$ representation
is discussed in \cite{RCKG00,levy18}.

\begin{figure}
\begin{center}
\includegraphics[width=0.8\textwidth,clip]{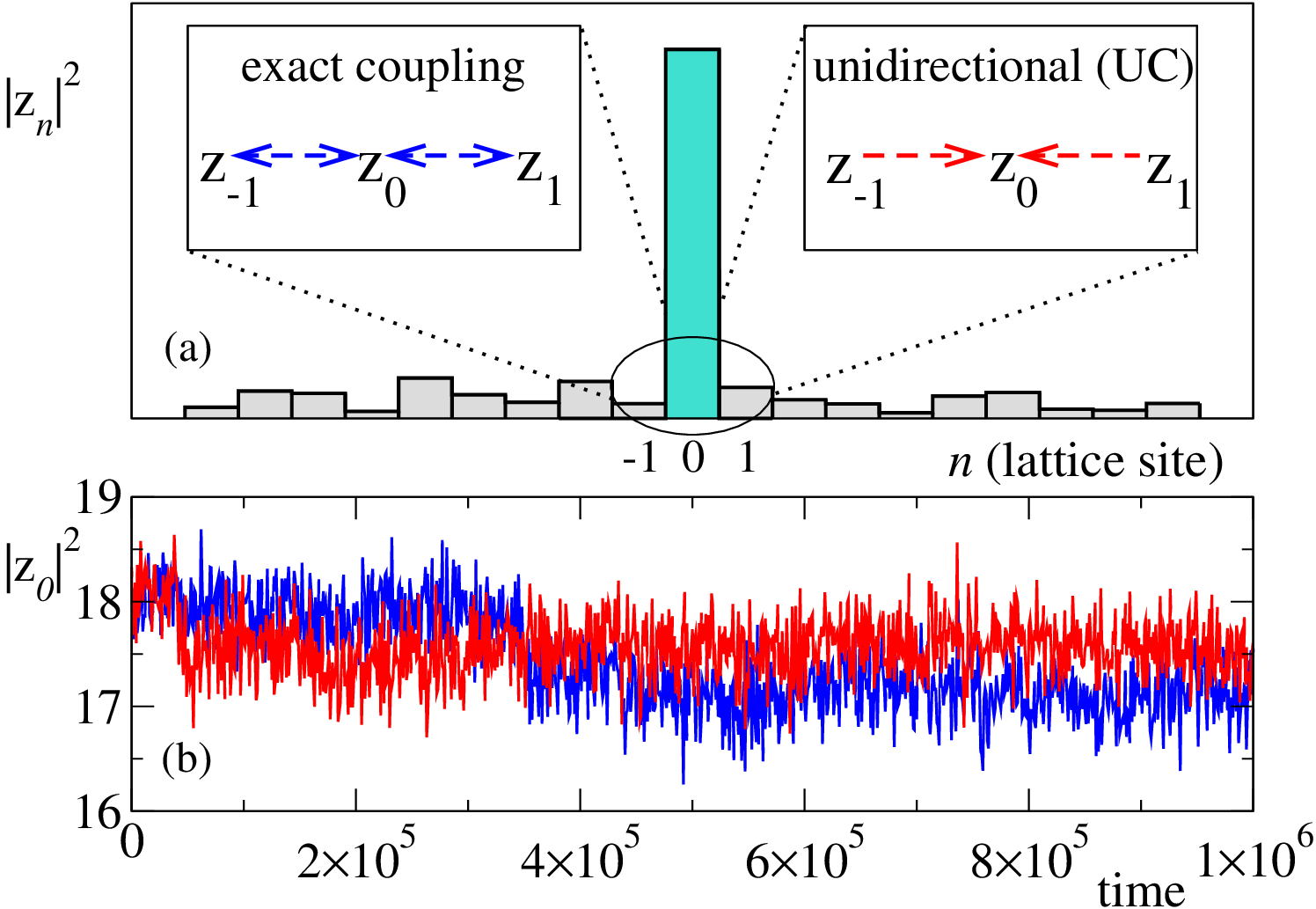}
\end{center}
\caption{
(a) A typical mass profile displaying a breather on site $n=0$ (cyan bar) and an incoherent background 
($n\ne 0$, grey bars). Left and right insets depict exact and unidirectional couplings between the breather site and the neighboring ones, respectively.
(b) Example of evolution of the mass of a breather with exact (blue line) and unidirectional coupling (red line) and
same initial conditions. Simulations are performed with $T=10$,  $\mu=-6.4$ and  $N_0=15$.
}
\label{fig.uni}
\end{figure}

In this paper we study the unidirectional coupling model, UC-DNLS, assuming that the breather placed on site $n=0$ feels the neighbouring
sites $n=\pm 1$, whereas the latter do not feel the breather, see Fig~\ref{fig.uni}(a). This amounts to modify Eq.~(\ref{eq:dnls}) in such a way that
the coupling term $z_{n-1}$ ($z_{n+1}$) is removed for $n=1$ ($n=-1$). More precisely, the evolution equations are
\be
\begin{array}{clcl}
i \dot {z}_0 &= -2 |z_0|^2z_0 - z_{1}-z_{-1} & (a) & n=0, \;\mbox{breather dynamics} \\
i \dot {z}_{\pm 1} &= -2 |z_{\pm 1}|^2z_{\pm 1} - z_{\pm 2} & (b) & n=\pm 1 \\
i \dot {z}_n &= -2 |z_n|^2z_n - z_{n+1}-z_{n-1} & (c) & |n| \ge 2
\end{array}
\label{eq.UC}
\ee

Accordingly, while the original DNLS model, Eq.~(\ref{eq:dnls}), is globally a Hamiltonian system, the UC-DNLS model
is composed of two subsystems that interact unidirectionally. 
%here is a a breather whose 
In detail, the breather dynamics is governed by a time-dependent Hamiltonian, see Eq.~(\ref{eq.UC}a), where 
$z_{\pm 1}(t)$ are external forcings.
On the other hand, the background, which has the standard DNLS form,  is decoupled from the site $n=0$
hosting the breather, see Eqs.~(\ref{eq.UC}b,c).
Altogether, this is a so-called master-slave configuration.

We can study the problem of breather relaxation in both full- and UC-DNLS, by assuming that
a breather of mass $b_0$ is initially present at site $n=0$, i.e. $|z_0(t=0)|^2 = b_0$, and that the
chain ends $n=\pm N_0$ are attached to thermal reservoirs.
 In this study we have implemented a single Langevin heat bath~\cite{iubini13} on site $n=N_0$ and assumed periodic
boundary conditions at the chain ends, i.e. $z_{N_0}=z_{-N_0}$. Hence, for given $(T,\mu)$-values, 
we have modified Eq.~(\ref{eq.UC}) for site $N_0$ as follows,
\begin{equation}
i \dot {z}_n = (1+i\gamma)\left[-2 |z_n|^2z_n - z_{n+1}-z_{n-1}\right] +i\gamma \mu z_n + \sqrt{\gamma T}\eta(t)\quad(n=N_0),
\end{equation}
where $\gamma$ is the coupling strength of the reservoir and $\eta(t)$ is a complex Gaussian white noise with zero mean and unit variance. Simulations were performed with $\gamma=1$ and the chain half-length ($N_0=15$) was chosen 
to be sufficiently large to avoid spurious effects of the stochastic heat bath dynamics on the breather site
\footnote{As clarified in section~\ref{sec.necessary}, a direct interaction ($N_0=0$) of the breather with a stochastic signal, 
as for example the reservoir, destroys the freezing effect. In Ref.~\cite{ICOPP} it was verified that for $N_0\geq 9$ this effect is negligible and that breather lifetimes are
independent of $N_0$.}.
In the following we assume that 
the background is thermalized with a mass density $a\simeq 1$ and that $b_0\gg 1$. Notice that the finite value of
$a$ does not allow one to approximate the background as a harmonic chain. In other words,
the breather interacts with a fully nonlinear, chaotic DNLS system~\cite{iubini21csf}, whose dynamics is not analytically treatable.

In the UC model, the background activity can be safely integrated by using a standard time step, thus allowing for much faster simulation
times: the only point that needs to be treated with care is the ``synchronization" of the background dynamics with that breather rotation,
which requires smaller time steps. The numerical approach is discussed in~\ref{app.num}.

Before discussing the model, we wish to show that it
is indeed able to reproduce at least qualitatively the freezing observed
for large breathers in the original DNLS model.
Two time traces of the breather mass corresponding to simulations of the DNLS equation (blue line) 
and of the UC equations (red line) are reported in
Fig.~\ref{fig.uni}(b) for the same thermodynamic configuration of the background. Given the relatively large time spanned 
in this figure, one has 
the qualitative impression that the feedback of the breather towards the
background  is negligible in this regime.
Strictly speaking, we must expect differences. For instance, within the UC model, the breather is not allowed to jump on the neighbouring sites,
whereas breather jumps can occur in the original DNLS model 
when the background amplitude is occasionally comparable to that of the breather itself.
The sporadic formation of ``dimers'' (i.e. transient bound states where the breather is delocalized over two lattice sites)~\cite{kenkre86} is another phenomenon that cannot emerge in the UC setup.
Nevertheless, it was already  argued~\cite{ICOPP}  that both phenomena become super-exponentially rare for increasing $b_0$,
therefore they are not the most important ingredients of the breather relaxation dynamics.
In this sense using the UC model even has the advantage to get rid of processes (jumps and dimer formations)
that are asymptotically irrelevant but that can mix with other, more relevant processes when the system size is finite.
Further considerations on the correspondence between the UC and DNLS are discussed in the conclusions.

\section{Necessary ingredients for frozen dynamics}
\label{sec.necessary}

In the UC model the breather is a nonlinear rotator (its frequency depends on its mass) which
feels the external signal $w(t) = z_{-1}(t) + z_{+1}(t)$, where $z_{\mp 1}(t)$ is the background signal 
at fixed temperature $0 < T < \infty$.
To better understand the role of breather nonlinearity, we discuss a simplified version of the problem in Eq.~(\ref{eq.UC})
in which the breather dynamics is linearized 
\be
\dot z = i\oz z + w(t),
\ee
where $\oz = -2 |z_0|^2$ represents the frequency of the breather, now assumed to be constant and determined by
the initial condition for  the breather mass $|z_0|^2=|z(t=0)|^2$.
The solution for $z(t)$ is trivial,
\be
z(t) = z(0) e^{i\oz t} + \int_0^t dt'e^{i\oz (t-t')} w(t') \; ,
\ee
from which we obtain the time dependence of the mass,
\bea
m(t) = |z(t)|^2  &=& |z(0)|^2 + \int_0^t dt'\int_0^t dt'' e^{i\oz (t'' -t')} w(t') w^*(t'') + \nonumber \\
&+&z(0)  \int_0^t dt'e^{i\oz (t-t')} w(t') + \mbox{c.c.}
\eea
The quantity of interest is
\be
\Delta m(t) \equiv m(t) - m(0) 
\ee
which should be averaged over espite it neitherthe external signal $w(t)$.
Since $w(t)$ has no preferential orientation, $\langle w(t)\rangle =0$ and we obtain
\be
\langle \Delta m(t) \rangle = \int_0^t dt'\int_0^t dt'' e^{i\oz (t'' -t')}
C(t'' - t'),
\ee
where $C(t) \equiv \langle w(0)w^*(t)\rangle$.

Taking the time derivative of both sides, we obtain
\bea
\!\!\!\!\!\frac{\partial}{\partial t} \langle \Delta m(t) \rangle &=&
\int_0^t dt'' e^{i\oz (t''-t)} C(t''-t) + 
\int_0^t dt' e^{i\oz (t-t')} C(t-t') = \nonumber \\
&=& \int_{-t}^t d\tau e^{i\oz \tau} C(\tau) \quad \rightarrow\qquad S(-\oz) .
\eea
Therefore, for large times
\be \langle \Delta m(t) \rangle = S(-\oz) t .\ee  

In conclusion, in the linear rotator model, the breather mass feels
a drift whose coefficient is the component of the power spectrum
of the external signal at the breather frequency, $S(-\oz)=S(2|z_0|^2)$.
It is therefore useful to determine the power spectrum
of the background signal $w(t) = -z_1 - z_{-1}$ sampled at equilibrium in the absence of breathers.

\begin{figure}
\begin{center}
\includegraphics[width=0.45\textwidth,clip]{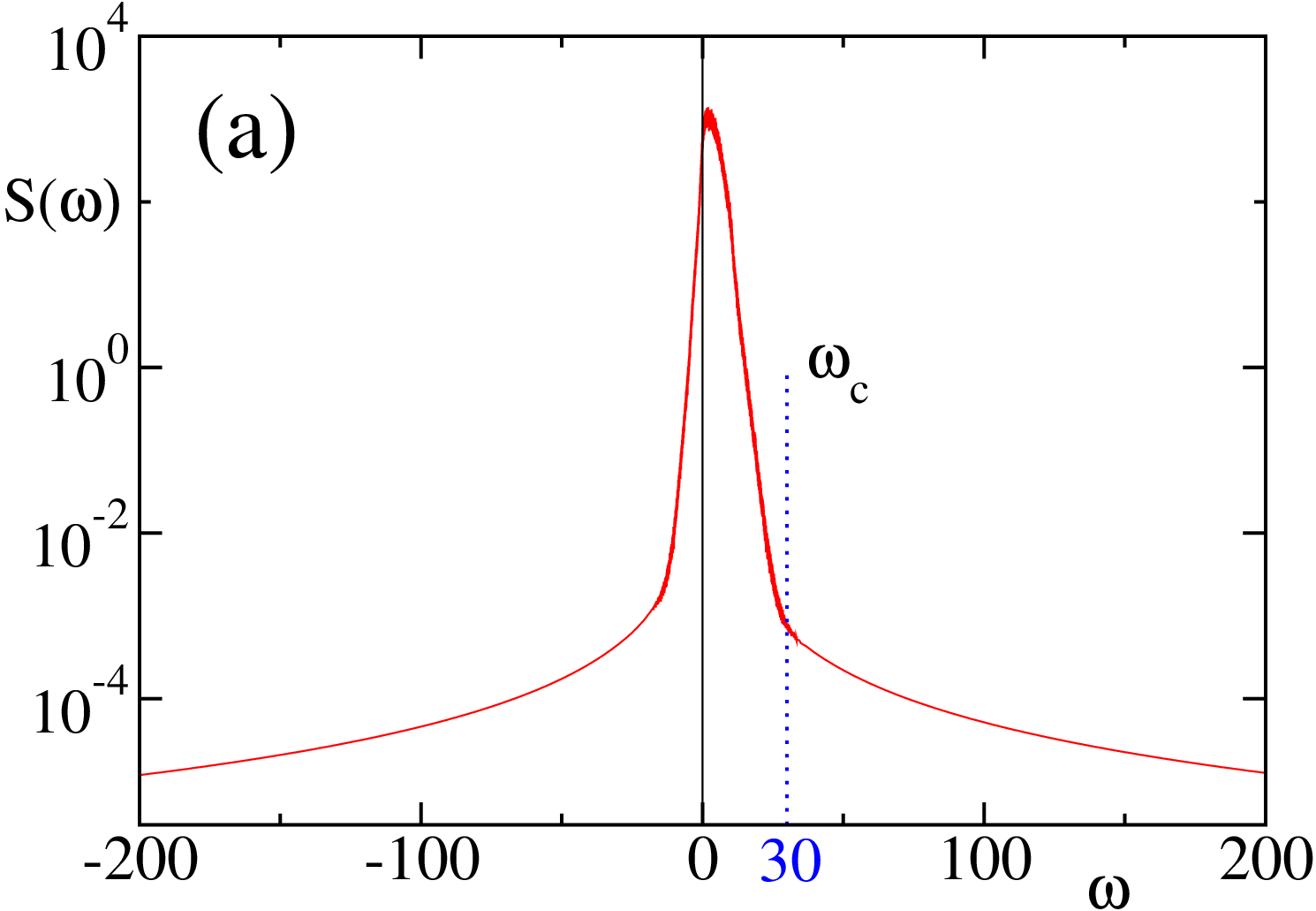}
\includegraphics[width=0.45\textwidth,clip]{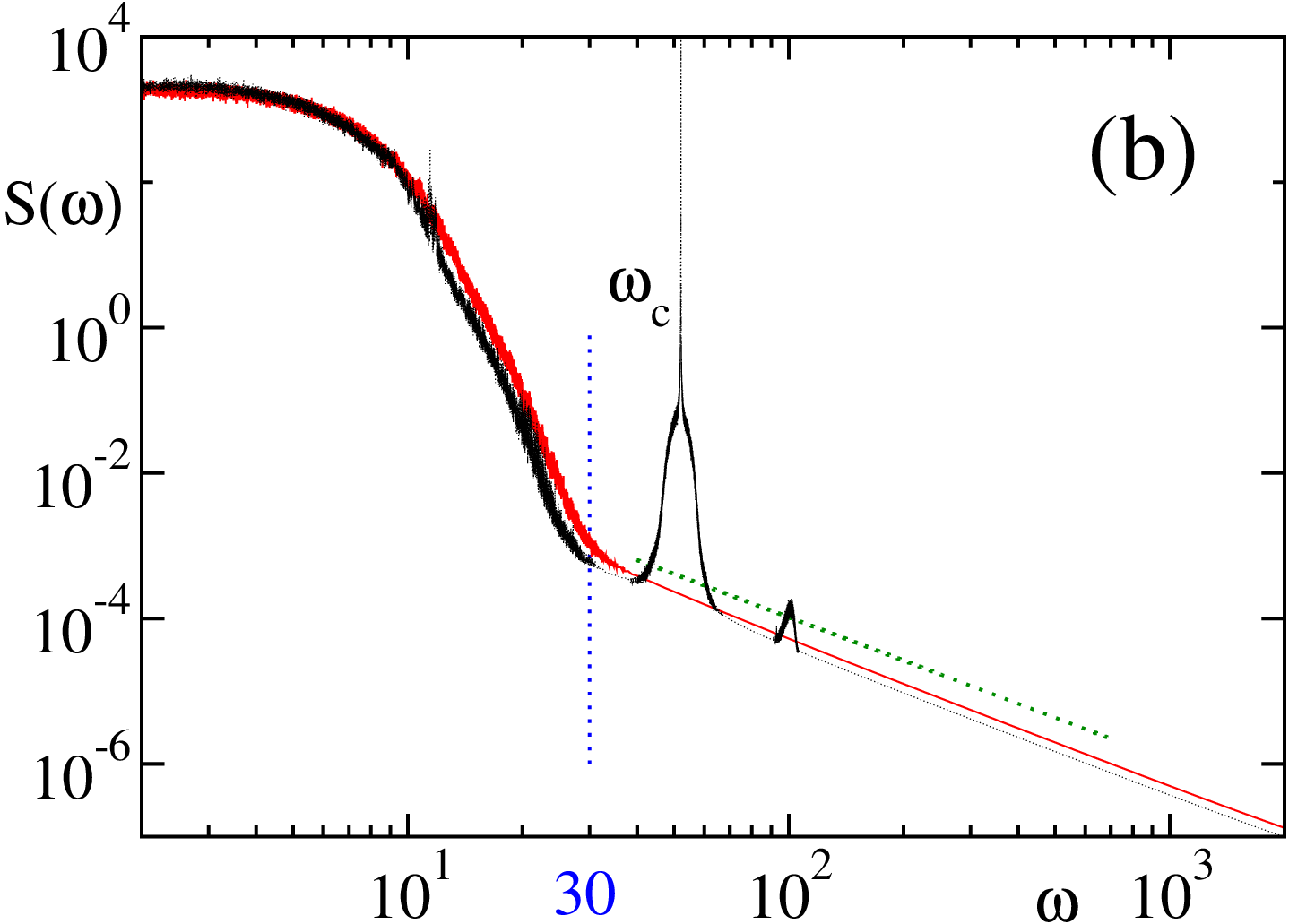}
\end{center}
\caption{
Power spectrum of $w(t) = z_1(t) + z_{-1}(t)$ for $T=10$ and $\mu=-6.4$. (a)~It decays exponentially for $\omega < \omega_c$.
(b)~For larger values of $\omega$ it decays as a power law, $S(\omega) \sim 1/\omega^2$, see the full red line.
The dotted black curve corresponds to the background spectrum for the full DNLS model, see the main text.
}
\label{fig.bgs}
\end{figure}

The spectrum for $T=10$ and $\mu=-6.4$ (these thermodynamical parameters set $a\simeq 1$ for the background chain) is shown in Fig.~\ref{fig.bgs}.
First of all we note that the spectrum is not symmetric, $S(\omega)\ne S(-\omega)$, because $z(t)$ is not real
and the single oscillators rotate on average anti-clockwise.
Second, and most importantly, 
$S(\omega)$ decays exponentially until a crossover frequency $\omega_c$ numerically found to be around $30$, then it decays as a power law with exponent equal to two.

This means that for large masses the dynamics of the linear model slows down as a power law 
and therefore this model cannot explain the observed frozen dynamics.
For the sake of openness, one should expect that in the original framework the background activity in the vicinity of the breather
may be affected by the breather itself (even though as a first approximation the breather can be replaced by a boundary condition
with an empty site).
To clarify this point, in Fig.~\ref{fig.bgs}(b) we have added the corresponding power spectrum for the full DNLS model (see the dotted black curve with spikes), 
which is, in fact, very close to the UC
one, except for a few localized features. The most valuable deviation is the peak at the breather frequency 
(the double of the breather
mass, set to be $b_0=26$). In spite of its height, its power is less than 1\% of the total power and so, essentially irrelevant.
Moreover, we cannot expect that an additional frequency component can slow down the breather dynamics. If it is going to play a
role is by facilitating the energy transfer.
A second yet minor deviation is the peak at twice the breather frequency: it is an expected harmonic, due to the nonlinearity
of the background dynamics.
Finally, notice the small peak at a lower frequency ($\approx 11$). It may be due to the sporadic formation of localized states
which involve the breather (e.g. dimers)~\cite{ICOPP}. Anyway, this is a minor deviation which does not modify the overall general scenario.
Altogether we can conclude that we cannot approximately treat the breather as a linear oscillator
 and it is thereby necessary
to reintroduce its nonlinear character as from Eq.~(\ref{eq.UC}) for $n=0$.

\begin{figure}
\begin{center}
\includegraphics[width=0.8\textwidth,clip]{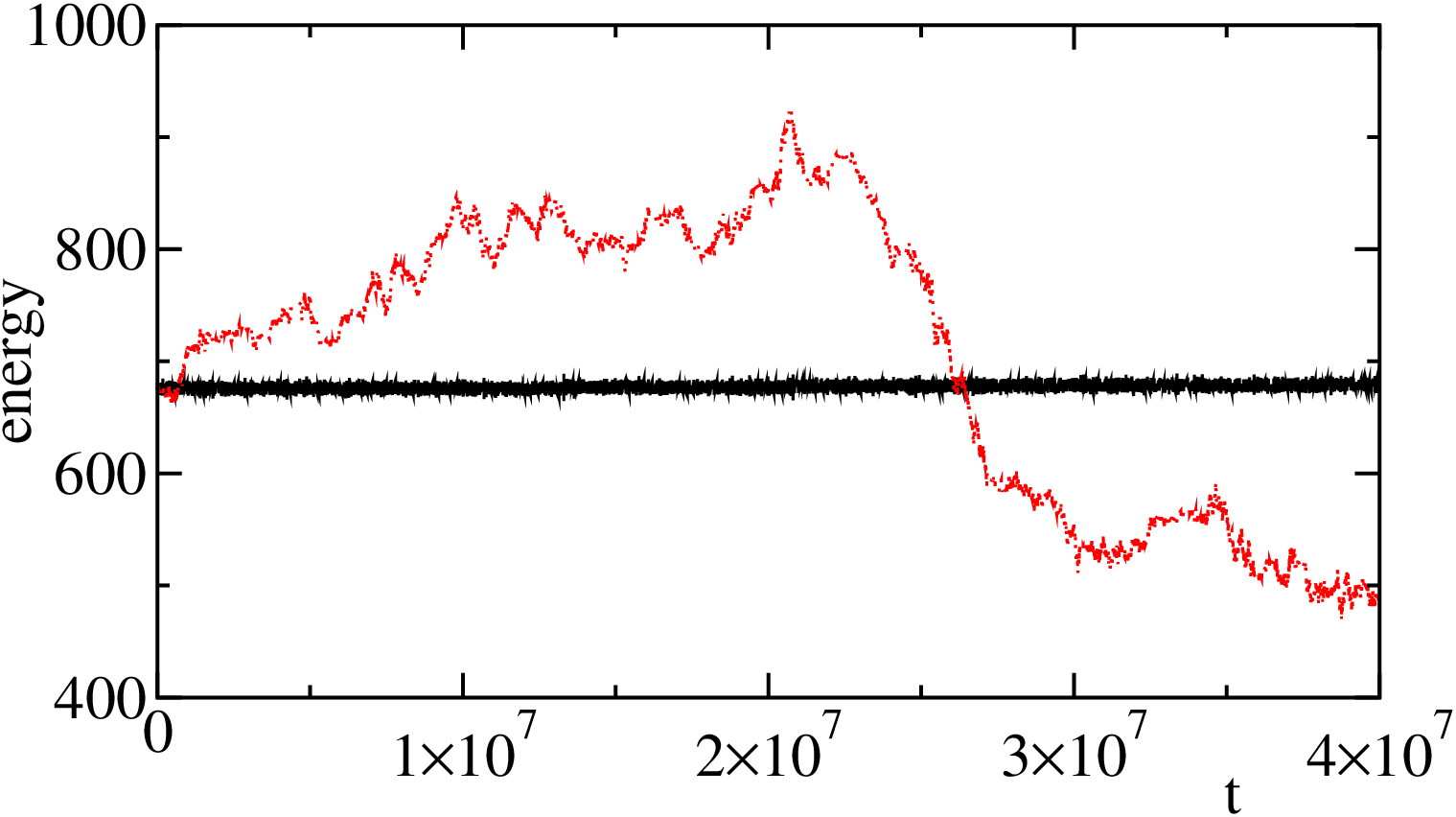}
\end{center}
\caption{
Time evolution of a breather of initial mass $b(0)=26$ and initial energy equal to $b^2(0)=676$ for the deterministic model (black full line)
Eqs.~(\ref{eq.UC}), and for the nonlinear stochastic model (red dotted line), see the main text.
}
\label{fig.sff}
\end{figure}

In order to identify those elements that are strictly necessary to induce a frozen dynamics, we considered yet
another approximate model by replacing the forcing term $w(t) = z_1(t) + z_{-1}(t)$ with a purely stochastic signal characterized
by the same power spectrum.
This has been done by computing the Fourier Transform of $w(t)$, randomizing 
the phases of the Fourier modes, and anti-transforming it.\footnote{More precisely, in order to generate very long signals, the transformation has
been computed separately in sequential intervals, and consecutive samples have been thereby reconnected to make them
sufficiently smooth.}
As shown in Fig.~\ref{fig.sff}, the breather dynamics resulting from  the stochastic model (see the red curve) is significantly more erratic than the deterministic one (black curve).
We can therefore conclude this section saying that both the nonlinearity of the breather site and the nonstochastic character of the background  are essential to induce frozen dynamics.

\section{Derivation of adiabatic invariants}
\label{sec.sufficient}

\subsection{Canonical adiabatic theory}

The main message of this subsection is that the dynamical evolution of the breather can be rewritten
into an Hamiltonian form, thus allowing for the application of 
the canonical perturbation theory and thereby deriving the first three orders of the AI.

The UC dynamics of the breather, see Eq.~(\ref{eq.UC}a), 
can be derived from the following time-dependent Hamiltonian, 
\be
{\cal H}_b = |z_0|^4 + z_0 (z_1^* + z_{-1}^*) + z_0^* (z_1 + z_{-1}) ,
\ee
where $z_{\pm 1}$ depend on time and represent the \textit{external signal} acting on the breather.
Using the canonical representations $z_0 = \sqrt{b} e^{i\theta}$ (the mass $b$ is $p-$type and the
phase $\theta$ is $q-$type), $z_{\pm 1} = (p_{\pm 1} + iq_{\pm 1})/\sqrt{2}$,
and defining the quantities $\lambda$ and $\alpha$ through the relations 
$\lambda \sin\alpha =(p_1 + p_{-1})$ and $\lambda\cos\alpha=(q_1 + q_{-1})$, we can rewrite
\be
{\cal H}_b = b^2 + \lambda(t)\sqrt{2b} \sin(\theta + \alpha(t)),
\label{eq.Hb}
\ee
where $\lambda$ and $\alpha$ encode the time dependence of the ``external" signal.

If we define the rescaled variable $J=b/b(0)\equiv b/b_0$, the new Hamiltonian is
$K=H_b/b_0$, i.e.
\be
K = b_0 \left[ J^2 + \frac{1}{b_0^{3/2}} \lambda \sqrt{2J}\sin(\theta+\alpha)\right] .
\ee
By now introducing the smallness parameter
 $\e \equiv 1/b_0$, we obtain
\be
K = \frac{1}{\e} \left[ J^2 + \e^{3/2} \lambda \sqrt{2J}\sin(\theta+\alpha)\right] \equiv 
\frac{1}{\e} H .
\ee
By then
rescaling time, $t_0 = t/\e$, we can write the Hamilton equations as
\be
\frac{\p\theta}{\p t_0} = \frac{\p H}{\p J}, \qquad
\frac{\p J}{\p t_0} = - \frac{\p H}{\p\theta} .
\ee 
For ease of notation and without risk of confusion, we redefine the time $t_0$ as $t$, so we finally have
to analyze the Hamiltonian
\bea
H &=& J^2 + \e^{3/2} \sqrt{2J} \lambda(\e t)\sin(\theta +\alpha(\e t)) \nonumber \\
&\equiv & H_0(J) + \e^{3/2} H_1(J,\theta,\e t) ,
\label{eq.H}
\eea
which has the standard form to be treated perturbatively.
Before proceeding, let us remind that the true breather Hamiltonian is $H_b = H/\epsilon^2 = b_0^2 H$ and that in
Eq.~(\ref{eq.H}) there are two time scales, $t$ and $\e t$: the former ($t$) corresponds to the breather rotation
and to the consequent oscillations of the breather mass (because the model is nonlinear);
the latter ($\e t$) is the ``old" time $t$ and corresponds to the dynamics of the background, therefore to the
forcing $\lambda(\e t)$ and $\alpha(\e t)$.

According to the canonical adiabatic theory~\cite{Lichtenberg_Lieberman},
the goal of the perturbative approach is to find a canonical transformation
\be
(J,\theta) \to (\bar J,\bar\theta)
\ee
such that the new Hamiltonian $\bar H$ is independent of $\bar\theta$ up to some order in $\e$.
The corresponding action variable $\bar J$ will be the sought after adiabatic invariant AI.
We should stress that the series defining the AI is asymptotic~\cite{Lichtenberg_Lieberman}. 
This means that for a given (initial) mass
of the breather, i.e. for fixed $\epsilon$, the series defining the AI improves up to a certain order $n^*$,
getting worse afterwards.\footnote{This is the way the physical system is telling us that an exact invariant 
exists only for $\e\to 0$. If the series was convergent for $\e < \e_c$ there would be a new 
conserved quantity for finite $\e$.}
The details of the calculation up to the second perturbative order are given in~\ref{app.cpt}.
Here we limit to sketch the method and to give the final expression.

Since $H_0$ does not depend on $\theta$, see Eq.~(\ref{eq.H}), at zero order $\bar J = J$,
and we need to use a canonical transformation which reduces to the identical transformation for $\e =0$.
Therefore, the starting point is the generating function of such identical transformation, 
$S_I (\bar J,\theta) =\bar J\theta$.
As shown in~\ref{app.cpt}, once we restore all the physical quantities, 
the final expression of AI at the second perturbative order is
\be
b_0 \bar J =
b + \lambda\frac{1}{\sqrt{2b}}\sin(\theta+\alpha) + \frac{1}{(2b)^{3/2}} \frac{\p}{\p t}
\left[ \lambda(t) \cos (\theta+\alpha(t))\right] .
\label{eq.cAI}
\ee

It is useful and also preparatory to the next section to take the square of the AI, obtaining
\be
Q\equiv (b_0 \bar J)^2 = b^2 + \lambda\sqrt{2b}\sin(\theta +\alpha) + \frac{1}{\sqrt{2b}} \frac{\p}{\p t}
\left[ \lambda(t) \cos (\theta+\alpha(t))\right] + o\left(\frac{1}{\sqrt{b}}\right) .
\label{eq.cAIs}
\ee
\commento{Since in Eq.~(\ref{eq.H}) the small parameter $\e$ appears with the different exponents
$\frac{3}{2}$ and $1$, a generic term of the expansion will be of order $\e^k$ with
$k=(3/2)n_1 + n_2$, with $n_{1,2}$ positive integers.} 
To avoid complicated notations we simply use the quantity $Q_k$ to mean $Q$ up to terms of
order $\e^k \sim b^{-k}$. So, for example, 
\bea 
Q_{-2} &=& b^2 \label{eq.Q-2}\\
Q_{-1/2} &=& Q_{-2} + \lambda\sqrt{2b}\sin(\theta +\alpha)\label{eq.Q-12}\\
Q_{1/2} &=& Q_{-1/2} + \frac{1}{\sqrt{2b}} \frac{\p}{\p t}
\left[ \lambda(t) \cos (\theta+\alpha(t))\right] \label{eq.Q12}.
\eea
We remark that $Q_{-2}$ is the square of the mass, i.e. the self-energy of the breather,
while $Q_{-1/2}$ is the full energy of the breather,%
\footnote{This result was obtained in Ref.~\cite{ICOPP} for the full, bidirectional DNLS model with the 
caveat that in such case the term proportional to $\lambda$ was $1/2$ of this,
because in the full model the coupling between the breather and the neighbouring sites
appears twice.}
see Eq.~(\ref{eq.Hb}) for $H_b$.

\subsection{Energy flux estimation}

Despite the procedure to go beyond $k=1/2$ within canonical adiabatic theory is well defined, calculations quickly become arduous.
To overcome this problem,
 we implement here a more efficient method based on 
the derivation of better and better approximations of the energy flux at the breather site. 

It is natural to refer to an energy-like variable, since this is the physical dimension of
the square of the AI $\bar J$ (called $Q$).
Moreover, $Q_{-1/2} = H_b$ because when the parameters $\lambda,\alpha$ are constant
$H_b$ is trivially conserved, $dH_b/dt = 0$. 

The idea is to decompose the time derivative $\dot H_b$ into the sum of a leading term, 
expressed as the time derivative of a suitable observable $L$ and
a higher-order correction term,
\begin{equation}
\dot H_b = \dot L + R,
\end{equation}
so that $H_b-L$ fluctuates less than $H_b$ and it is a better approximation of the AI. 
One can then repeat this approach by decomposing $R$ and so on -- see~\cite{deroeck14} for 
a rigorous discussion of
analogous techniques.
This way, we obtain the various $Q_k$ approximants for increasing $k$,
which coincide with the canonical adiabatic expressions
(\ref{eq.Q-2},\ref{eq.Q12}) for $k\le \frac{1}{2}$.
The mathematical details of the method are illustrated in~\ref{app.apt},
here we limit to write explicitly the next order of $Q$,
\be
Q_{3/2} = Q_{1/2} +
\left[\frac{i}{4\sqrt{2}b^{3/2}}e^{i\theta} \frac{\p^2}{\p t^2}\left(\lambda e^{i\alpha}  \right)  +c.c.  \right] , 
\ee
because $Q_{3/2}$ will be used to produce most of the results of this paper.%
\footnote{Notice that the evolution equations of the background chain allow expressing the
time derivatives of the external signal $\partial_t^n  \lambda(t),\alpha(t)$ appearing in $Q_k$
in terms of the background variables
 $\left[z_{\pm 1}(t),z_{\pm 2}(t),\cdots,z_{\pm (n+1)}(t)\right]$ and of their complex conjugates.}

\begin{figure}[ht]
\begin{center}
\includegraphics[width=0.8\textwidth,clip]{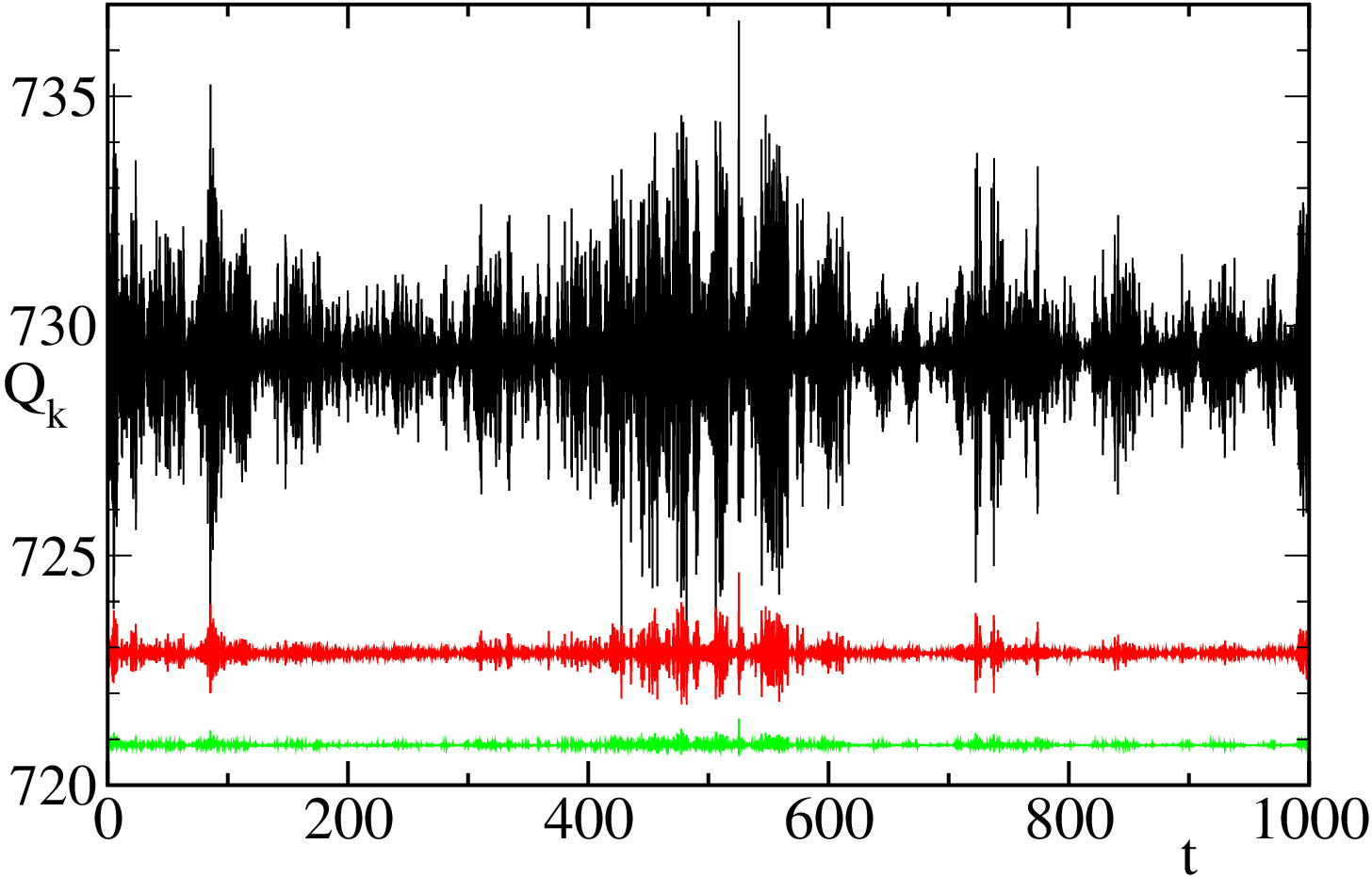}
\end{center}
\caption{
From top to bottom: $Q_{-1/2}(t)$, $Q_{1/2}(t)$, and $Q_{3/2}(t)$ ($Q_{-2}$ is not
plotted because its fluctuations are too large). $Q_{1/2}(t)$ and $Q_{3/2}(t)$ have been
vertically shifted for clarity. The time evolution refer to a breather of initial mass $b(0)=27$.
Simulations correspond to $T=10$ and  $\mu=-6.4$.
}
\label{fig.H0123}
\end{figure}

As a preliminary, qualitative result, in Fig.~\ref{fig.H0123} we compare the time fluctuations of 
$Q_{-1/2},Q_{1/2}$, and $Q_{3/2}$
for a breather with initial mass $b_0=27$ and a background chain with $T=10$. Clearly, despite the presence of the 
incoherent background dynamics, we obtain a manifest reduction of fluctuations for increasing orders of approximation of $Q_k$.

\section{Quantifying frozen dynamics}
\label{sec.results}

\begin{figure}
\begin{center}
\includegraphics[width=0.8\textwidth,clip]{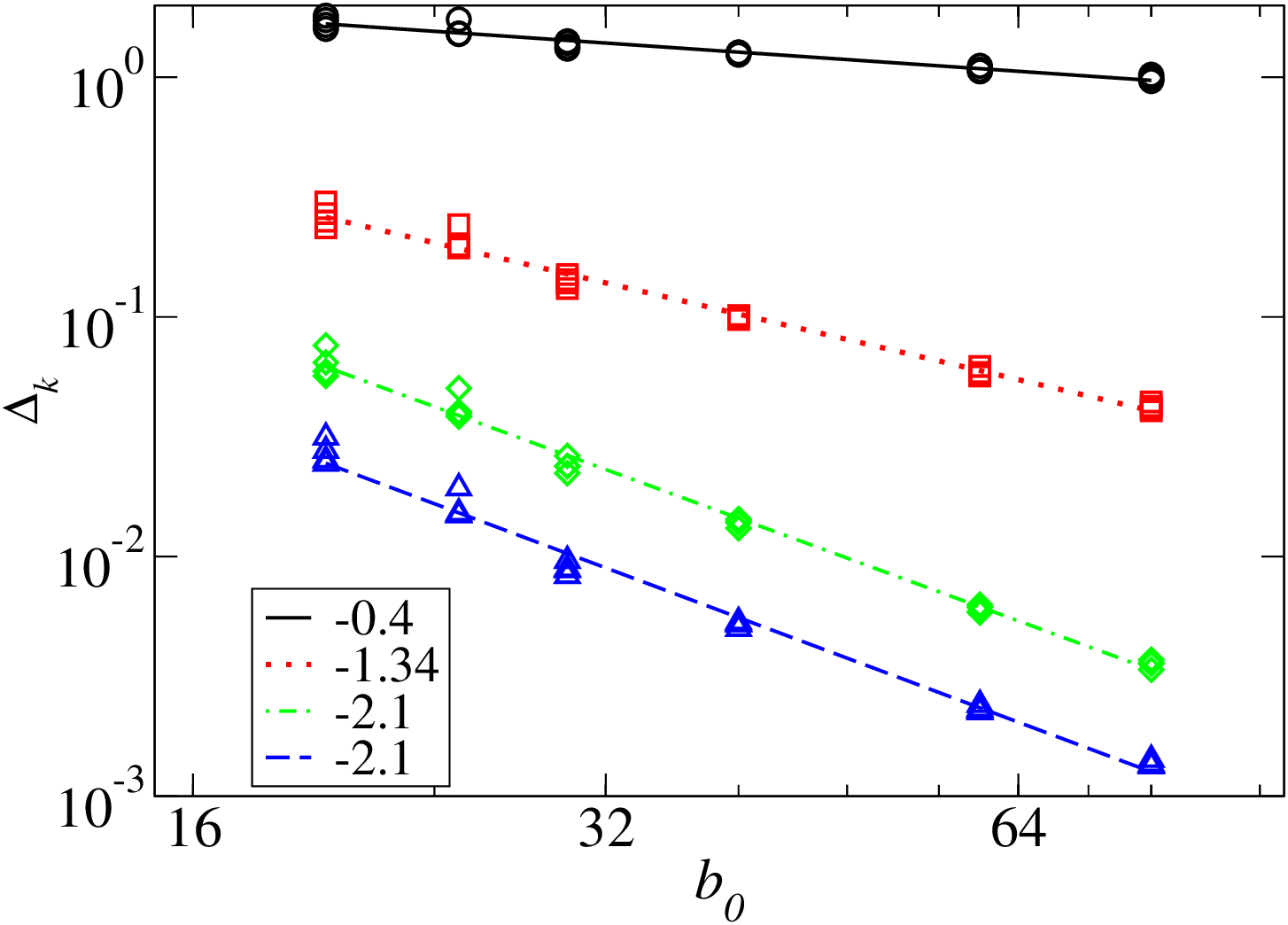}
\end{center}
\caption{
The maximal variation of $Q_k(t)$ over a time interval $b_0$ {\it versus} $b_0$, scaled with respect to the average value of the
forcing signal (see the main text). From top to bottom: $k=-1/2,1/2,3/2,5/2$. 
Each point is an average over 600 temporal blocks, sampled any $dt=0.4$.
Dashed lines are best fits whose slopes are given in the legend.
}
\label{fig.sH0123}
\end{figure}

A quantitative analysis of the scaling of the fluctuations of $Q_k$ with the breather mass $b_0$ 
is presented in
Fig.~\ref{fig.sH0123},
where we plot the maximal variation of $Q_k(t)$ over a time interval equal to the ratio between slow and fast scales~\cite{Lochak_Meunier}, namely $1/\e = b_0$.
More precisely, we plot
\begin{equation}
\Delta_k \equiv \left \langle
\frac{ \mbox{max}_{t\in (t_0,t_0+b_0)} |Q_k(t)-Q_k(t_0)|}{|\overline w|} \right \rangle
\end{equation}
where the angular brackets denote an average over different realizations, while the overline in the denominator means an average
over the time interval $[t_0,t_0+b_0]$.

$\Delta_k$ appears to decrease as $b_0^{-\chi}$ where $\chi$ is in principle the order of the successive term to $k$ 
in the perturbative expansion. This is approximately true for $\Delta_{-1/2}$ ($\chi=0.4$) and
for $\Delta_{1/2}$ ($\chi=1.34$). 
It is no longer true for
$\Delta_{5/2}$ which, although being
smaller than $\Delta_{3/2}$, decreases with the same rate, if not slower. We comment more on this issue in the following section.

\begin{figure}
\begin{center}
\includegraphics[width=0.8\textwidth,clip]{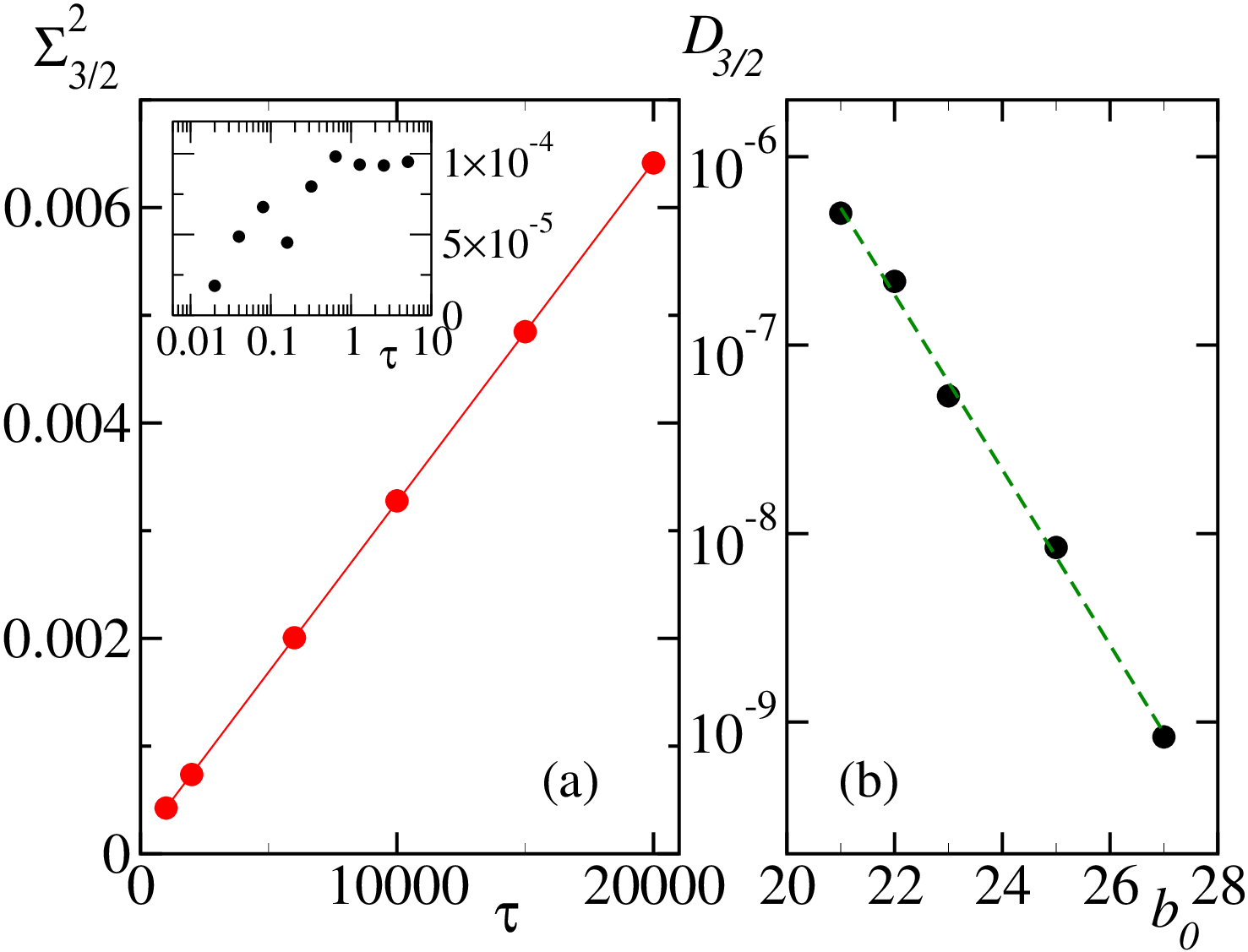}
\end{center}
\caption{
(a) The variance of the fluctuations of $Q_{3/2}$ over an increasing time interval $\tau$, for $b_0 = 22$.
Inset:  detail for $\tau \lesssim 10$ (notice that the horizontal scale is logarithmic). The variance initially increases to attain an approximately constant value.
Main: $10 \ll \tau \le 20000$. The variance has a neat linear increase
(the solid line is a fit) which allows to define a precise diffusion coefficient $D_{3/2}(b_0)$ through
the relation $\Sigma_{3/2}^2(\tau) = 2D\tau$.
(b) The diffusion coefficient $D_{3/2}(b_0)$, derived for different initial masses
of the breather. The fitted slope is -1.07.
}
\label{fig.dd}
\end{figure}

Once clarified the behavior of $Q_k$'s for increasing $k$,
 we can focus
on their long-term fluctuations to determine whether and how they grow in time. 
In the following we will make reference to $Q_{3/2}$, which provides an accurate approximation of the adiabatic
invariant for the
breather masses investigated in this paper and is computationally less expensive than $Q_{5/2}$. 

Let $\delta Q_{3/2}(t,\tau)=Q_{3/2}(t+\tau)-Q_{3/2}(t)$ denote the variation of $Q_{3/2}$ from $t$ to $t+\tau$.
Upon assuming that the evolution of $Q_{3/2}(t)$ is a stationary process (this point will be discussed later),
we consider the second moment 
\begin{equation}
\Sigma^2_{3/2}(\tau) = \overline {\left(\delta Q_{3/2}(t,\tau)\right)^2}= \overline{(Q_{3/2}(t+\tau)-Q_{3/2}(t))^2 }\,,
\end{equation}
where the overbar denotes again an average over time $t$.
In practice, $\Sigma^2_{3/2}(\tau)$ is
 a measure of the accuracy of our approximate AI expression as a function of time. 
Due to the absence of exact additional conservation laws arising from the breather dynamics, one expects 
$\Sigma_{3/2}^2(\tau)$ to
grow with $\tau$, as shown in Fig.~\ref{fig.dd}(a) from simulations performed for $b_0=22$. 
More precisely, after an initial transient, where $\Sigma_{3/2}^2(\tau)$ rapidly increases from zero towards an approximately
constant value (see the inset), we have a diffusive regime, testified by
a clean linear behavior as reported in the main panel~(a).
We can therefore extract the diffusion coefficient $D_{3/2}=\Sigma_{3/2}^2(\tau)/(2 \tau)$ and thereby  investigate its dependence on the breather mass.
The results are reported in Fig.~\ref{fig.dd}(b), where we see that the
diffusion coefficient decreases exponentially with $b_0$, 
$D_{2} \simeq D^* \exp(-c b_0)$, with $c\simeq 1.07$.
From a numerical point of view this is the most convincing evidence of the frozen dynamics of the UC model.

\begin{figure}
\begin{center}
\includegraphics[width=0.8\textwidth,clip]{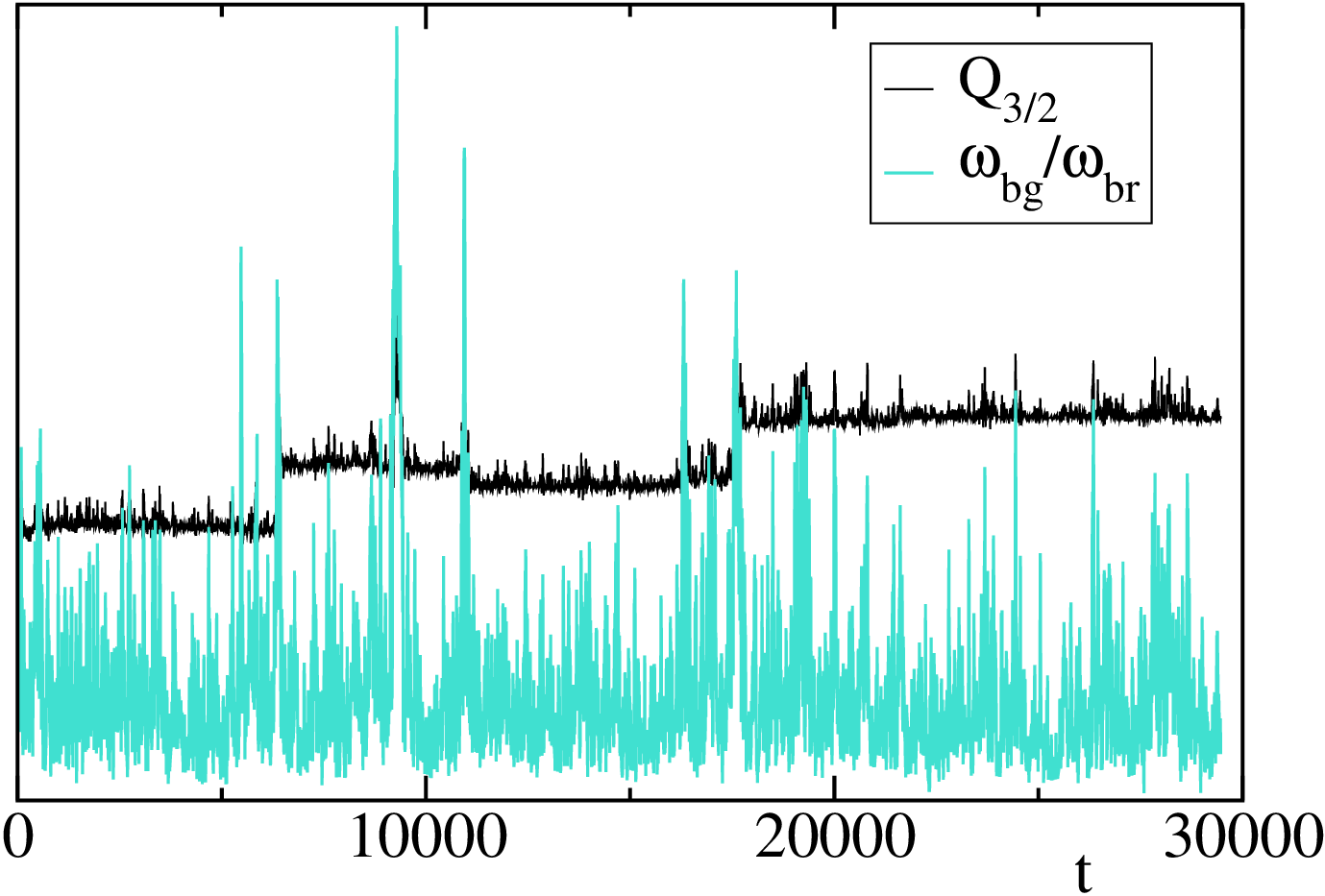}
\end{center}
\caption{
Upper curve (black): the quantity $Q_{3/2}(t)$.
Lower curve (cyan): the ratio $\omega_{bg}/\omega_{br} = |w(t)|^2/|b(t)|^2$.
The initial mass of the breather is $b(0)=27$.
}
\label{fig.jH3}
\end{figure}

In order to shed some light on the underlying diffusion mechanism, in
Fig.~\ref{fig.jH3} we plot $Q_{3/2}(t)$ over a time interval much longer than in Fig.~\ref{fig.H0123}
(see the upper black curve).
The fluctuations are strongly asymmetric and characterized by several spikes, but the most relevant
feature is the presence of jumps, which are the main source of the diffusion process.
A priori, one might qualitatively expect the jumps to be associated with occasional large $w$-values,
since these bursts likely have a stronger impact onto the breather amplitude.
It is not however easy to transform this natural feeling into something quantitative.
The best proxy we have found is the ratio between the background and the breather instantaneous
frequency $\omega_{bg}/\omega_{br} = |w(t)|^2/|b(t)|^2$ 
%where $\omega_{bg}$ is determined from the rotation of $w(t)$  
(see the lower curve in Fig.~\ref{fig.jH3}, suitably scaled to allow for a 
clear comparison with the behavior of $Q_{3/2}$).
The comparison confirms the intuition that jumps and spikes are correlated with the
occurrence of large fluctuations in the background.
Moreover, this explains the asymmetry: only a large background can reduce significantly the separation
of time scales required for the existence of an AI.

\begin{figure}
\begin{center}
\includegraphics[width=0.8\textwidth,clip]{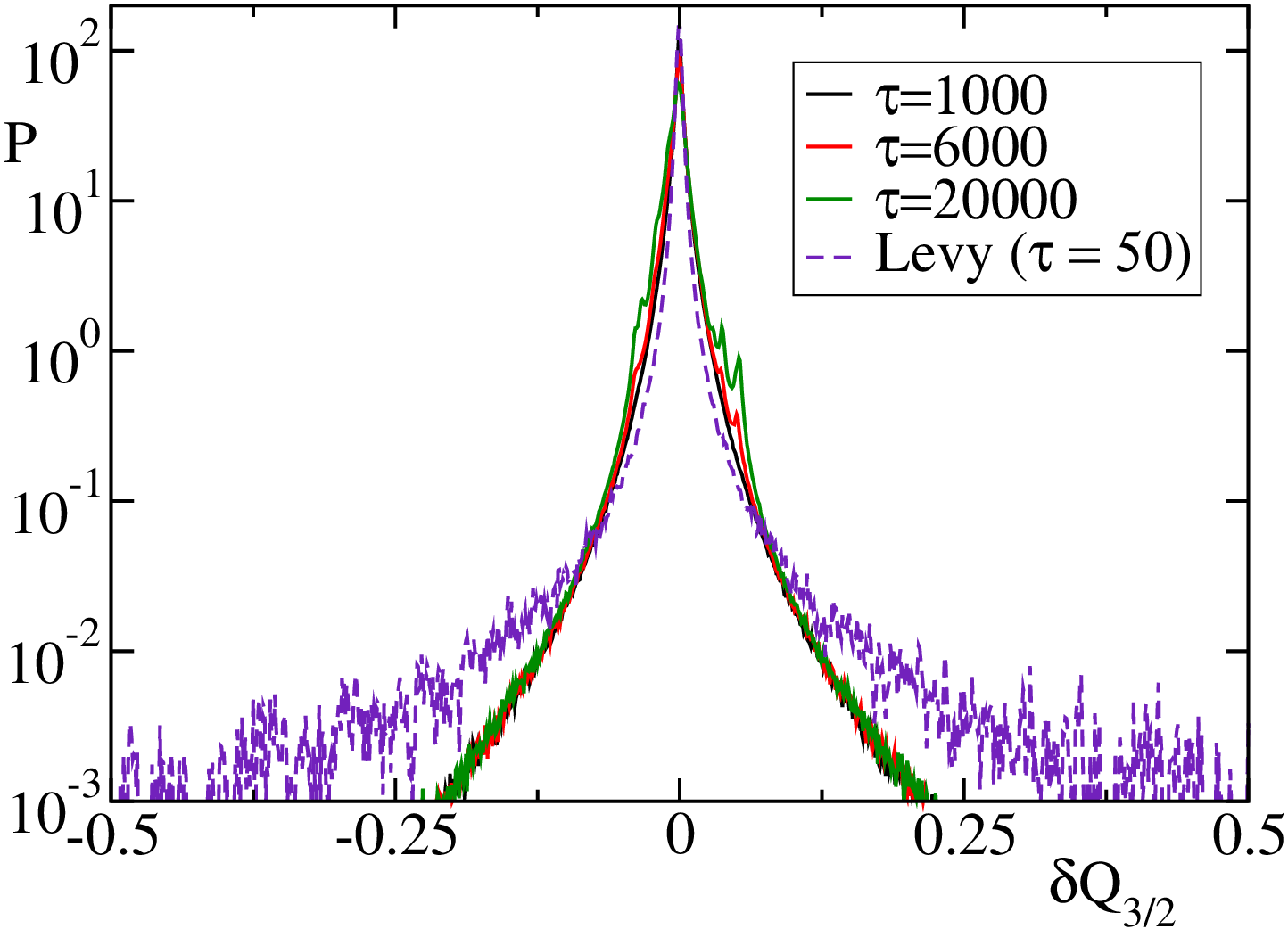}
\end{center}
\caption{
The probability distribution function $P(\delta Q_{3/2})$ for the variation of $Q_{3/2}$ over a time $\tau$
for three different values of $\tau$
(full lines, see legend). 
Dashed line: probability distribution function for $\delta x \equiv x_{n+1} - x_n$ and $\alpha=2.2$, see
the main text.
}
\label{fig.ef}
\end{figure}

A more complete characterization of the diffusion process is obtained by looking at the probability distribution of
$\delta Q_{3/2}(\tau)$. Three distributions (properly rescaled to the same width)
detected at different times are presented in Fig.~\ref{fig.ef}. 
The average is statistically indistinguishable from zero, suggesting that if a drift is present, it is
negligible.

Instead of resembling Gaussians (as expected
for a genuine diffusive process), they resemble more L\'evy distributions.
In fact, by looking at the  evolution of $Q_{3/2}$ in Fig.~\ref{fig.jH3}, it is natural to see it as a series of
jumps followed by steady quiescient periods. Accordingly, one can think of the following schematization:
a discrete-time stochastic process 
\begin{equation}
x_{n+1} = x_n + \xi_n
\end{equation}
where $\xi_n$ is a random variable distributed as $\xi^{-\alpha}$ in the range $[1,\infty)$.
The resulting distribution for $\alpha=2.2$ is presented in Fig.~\ref{fig.ef}: it overlaps very well
with the curves obtained from the UC model over three decades.

Without pretending that such a toy model exemplifies
the diffusion of the AI, we can nevertheless observe that the exponent $\alpha$ being larger than 2 implies
that a standard diffusion is eventually expected (as indeed suggested by the direct simulations).
The deviations from a Gaussian are simply finite-size effect: much longer times would be required to
see a normal distribution. This slow convergence represents a further obstacle to a more accurate comparison
with L\'evy-like processes.

\section{Discussion and open problems}
\label{sec.disc}

The presence of a breather in the system induces the existence of two distinct time scales in the problem:
(i) a slow time scale of order one for the dynamics of the background; (ii)
a fast time scale of order $1/b$, i.e. the rotation period of a breather of mass $b$.
The ratio between these time scales defines a smallness parameter $\e =1/b$, which allows to approach the relaxation dynamics
perturbatively. In Ref.~\cite{ICOPP} we considered the full DNLS Hamiltonian $H$ and searched for an approximate
constant of motion $Q$, imposing that their Poisson bracket vanishes, $\{Q,H\}=0$.
However, we were unable to go beyond the first perturbative order at which $Q$ coincides with the breather energy $H_b$.
For this reason,
in this paper we have approached the problem with a simplified unidirectional model, where the interpretation of slow dynamics
in terms of an AI is more transparent. In this model, the background has
its own dynamics, and it does not feel the breather, while the latter feels the dynamics of the background
which acts as an external signal. The AI comes into play when otherwise constant parameters acquire a slow time
dependence. 
In our case, the ``constant parameters" are the $z_{\pm 1}$-amplitudes in the breather neighbouring sites.
Meanwhile, the breather is a nonlinear rotator whose rotational frequency equals two times its mass. 

If $z_{\pm 1}$ are constant, the breather mass oscillates periodically at the same high
frequency of its rotation.
If $z_{\pm 1}$ slowly depend on time (slowly with respect to the intrinsic frequency of the breather)
it is natural to conjecture that an AI exists much in the same way it exists either 
for a pendulum with slowly varying length~\cite{Goldstein}, 
or for a free particle in a box of slowly varying size~\cite{Lochak_Meunier}.
Here, with the help of canonical perturbation theory, we have been able to determine different approximations
of the AI which, as usual, can be identified with a suitable action variable.
Equivalent results have been obtained by following a perhaps more intuitive approach, based on the
analysis of the energy flux.
The very existence of an AI provides a solid foundation for a full understanding of
the frozen breather dynamics in the DNLS equation.

An exact AI cannot exist for a finite $\e$, as this would imply the existence
of stationary breathers over infinite time scales. Mathematically, this is a consequence of the 
asymptotic nature of the  perturbative expansion of the AI, i.e. the fact that
the expression of $Q$ (the AI) can be improved only up to a certain order $n^*$:
going beyond $n=n^*$ makes $Q$ more fluctuating.
Our simulations (see Fig.~\ref{fig.sH0123})
show that the improvement obtained in
passing from $k=2$ to $k=5/2$ does not advance upon increasing $b$ (decreasing $\e$),
as if $n^*$ was independent of $\e$. 
This result sounds anomalous, since it suggests that the ``quality" of the expansion does not improve
when the breather amplitude increases. 
Performing simulations for significantly larger values of the breather mass might clarify this
point but they would require exceedingly long computation times. 

A better understanding of the AI violation may come from a detailed identification of the
mechanisms responsible for the jumps observed in Fig.~\ref{fig.jH3}.
In our previous study of the full DNLS model~\cite{ICOPP} we identified a possible mechanism in the formation
of bound states such as dimers, induced by rare mass fluctuation in a site close to the breather.
Dimers are almost decoupled from the rest of the system, and
its two sites exchange periodically mass and energy. At a certain point the dimer breaks and may release energy towards the background.
However, as argued in~\cite{ICOPP}, the frequency of this phenomena is so small that
the breather lifetime would be super-exponential with its mass rather than simply exponential as numerically observed.
In the UC model, dimers cannot form at all (they need a bidirectional coupling between breather and background) and yet, we
have seen that the breather life-time is exponentially long, confirming the previous intuition that the basic
mechanisms should be searched for elsewhere.
Fig.~\ref{fig.jH3} suggests the AI jumps are due to anomalously large value of the frequency in the neighbouring sites.
The distribution of $\delta Q$ displayed in Fig.~\ref{fig.ef} reveals a similarity between the evolution
of the AI and a L\'evy process at the borderline between normal and anomalous diffusion. 
However, we have not been able to identify in a quantitative way the specific nature of the events which 
induce AI-variations; this is still an open problem.

Another  open issue concerns the drift of the breather mass.
In the full DNLS model, it is clear that in the thermal region ($h<2a^2$), the breather amplitude is expected to decrease,
because the equilibrium state does not present breathers at finite, positive temperature.
Once we pass to the UC model, the Hamiltonian is time-dependent, and breather relaxation is not necessarily
induced by general thermodynamic principles.
In the stochastic models discussed in~\cite{IPP14,IPP17} it is possible to prove 
that the breather is still absorbed in the case of unidirectional coupling.
For the deterministic UC-DNLS model herein studied numerics is not convincing. 
The average value of the distribution displayed in Fig.~\ref{fig.ef} is not distinguishable from zero.
Simulations performed for a smaller amplitude ($b=20$) suggest a weak negative drift (around $-2\times 10^{-6}$).
A more refined analysis is necessary.

Finally, among the possible open perspectives, it would be interesting to investigate whether a similar freezing mechanism
involving AI's arises also in other classes of  oscillators models, as the already mentioned Klein-Gordon model or the rotor chain.
Indeed, despite their different symmetries and the absence of an equilibrium condensed phase, these models can
sustain stable breather states for large enough energy densities, as in the DNLS equation.

\section{Acknowledgements}
PP acknowledges support from the MIUR PRIN 2017 project 201798CZLJ.

\appendix

\section{Integration algorithm}
\label{app.num}

In this Appendix we present the optimized integration method employed for  numerical evolution of breathers
with very large mass in the UC model. 
In this limit, standard integration routines such as the Runge-Kutta algorithm
become quickly very inefficient due to the large separation of time scales between breather and backgrounds.
On the one hand, the integration of the breather site would require a very small time step $dt_b$ in order
to sample accurately the fast dynamics. On the other hand, the background dynamics is essentially frozen over this timescale.

 To overcome this problem,  we derive 
 here a specific integration scheme which exploits the unidirectional nature of the breather-background interaction.
  The background is evolved with a time step $dt$
 suitably chosen to ensure a desired accuracy of the integration over the background degrees of freedom.  For the regimes
 explored in this paper, one has $dt \gg dt_b$. The unidirectional breather evolution is thereby implemented as a suitable symplectic integration.

More precisely, 
we perform a piecewise constant approximation of the external background signal $w(t)$,
 \begin{equation}
 w(t)=w_n  \quad n dt\leq t < (n+1) dt,\, n\in \mathbb{N}
 \end{equation}  
where $w_n=w(n dt)$ is the generated external signal.

Within each  interval $T_n=[n dt, (n+1) dt]$ the equations of motion of the breather with unidirectional coupling read
\be
\label{eqbr}
i\dot z_0=-2|z_0|^2 z_0 + w_n \; .
\ee
Since $w_n$ is constant in $T_n$, we can introduce the Hamiltonian 
\be
H=|z_0|^4 + z_0 w_n^* + z_0^* w_n\,,
\ee
which generates Eq.~(\ref{eqbr}) limited to $T_n$ through the Hamilton equations $\dot z_0 = -\partial H/ \partial (iz_0^*) $.
A symplectic integration scheme over the interval $T_n$ can be now introduced according to Ref.~\cite{boreux2010}. In detail, we decompose
$H= H_1 + H_2	$, where $H_1= |z_0|^4$ and $H_2= z_0 w_n^* + z_0^* w_n$.  The equation of motion generated separately by $H_1$  reads
\be
\label{eq:eqh1}
i\dot z_0=-2|z_0|^2 z_0
\ee
and corresponds to a pure rotation of $z_0$ in the complex plane 
\be
\label{eq:solh1}
 z_0(t)=z_0(0) \exp(i\omega_0  t)\, ,
\ee
where  $\omega_0=2|z_0(0)|^2$ is the breather frequency. Notice that $\omega_0$ is a conserved quantity of
the dynamics in Eq.~(\ref{eq:eqh1}).
 The evolution generated by $H_2$
reads
\be
i \dot z_0=w_n
\ee
and identifies the translation
\be
\label{eq:solh2}
 z_0(t) = z_0(0) -i w_n t\,.
\ee 
 Denoting formally by $\exp(H_1 t)$ and $\exp(H_2 t)$ the two propagators 
associated respectively to $H_1$ and $H_2$, the   simplest (second order) integration scheme reads

\be
P(t)=\exp(H_1 t /2) \exp(H_2 t) \exp(H_1 t /2)\,,
\ee
where $P(t)$ is the overall propagator of the breather degrees of freedom.

Altogether, the integration scheme in the time interval $T_n$  amounts to the following steps:
\begin{enumerate}
\item Consider the initial state $z_0(n dt)$ and external forcing $w_n$; 
\item  compute $\omega_0=2|z_0(n dt)|^2$ and perform a pure rotation
by an angle $\omega_0 dt/2$  according to Eq.~(\ref{eq:solh1});
\item update $z_0 $ according to Eq.~(\ref{eq:solh2});
\item update $\omega_0$ and perform another pure rotation by $\omega_0 dt/2$.
\end{enumerate}

At the end of this procedure, one updates the external forcing  and starts again with $n\rightarrow n+1$. 
In our simulations we have adopted the above scheme and verified that for the parameters considered in this paper 
$dt=10^{-5}$ allows to reach sufficient numerical precision. We finally remark that the precision
of the breather dynamics can be further improved by implementing higher order symplectic schemes as in~\cite{boreux2010}. 
The overall computational cost of such an improvement 
usually turns out to be modest, as it affects the breather site only.

\commento{   % Beginning of Comment
\section{Summary of the theory of Adiabatic Invariants}
\label{app.ai}

In this article we use the canonical adiabatic theory as expounded by Lichtenberg and Lieberman (LL)~\cite{Lichtenberg_Lieberman}.
We have also used the book by Lochak and Meunier (LM)~\cite{Lochak_Meunier} and the Encyclopedia edited by Scott (SE)~\cite{Scott_Encyclopedia}. 

According to SE, if we have an Hamiltonian system depending on time dependent parameters $\lambda(t)$ and $T$ 
is the ratio between the slow and the fast time scale, the constancy of the adiabatic invariant depend on the analytic character
of the forcing. If $\lambda(t)$ is ${\cal C}^\infty$ (resp. ${\cal C}^m$) the accuracy of preservation is exponentially good, i.e. the change 
of the adiabatic invariant is
\be
\Delta I = \alpha \exp(-\beta T) , \quad \mbox{resp.} \quad
\Delta I = \alpha T^{-(m+1)} .
\ee
Please, note that in these expressions it is not clarified over with time scale $\Delta I$ is evaluated.

In LM authors work with an Hamiltonian $H(p,q,\tau)$ which is ${\cal C}^\infty$ in the canonical variables $(p,q)$
and of class ${\cal C}^{k+1}$ in the slow time variable $\tau=\e t$ ($0 \le \tau \le 1$).
Then, there exists a $k-$th order invariant $I_k(p,q)$ such that \textit{for any trajectory} one has
\be
|I_0 - I_k| = O(\e), \qquad
\sup_{t\in [0,1/\e]} |I_k(t) - I_k(0)| \le c_k \e^{k+1} .
\ee
} % End of Comment

\section{Details of the canonical perturbative approach}
\label{app.cpt}

We start from the breather Hamiltonian given in Eq.~(\ref{eq.H}) and reproduced
here for completeness,
\bea
H &=& J^2 + \e^{3/2} \sqrt{2J} \lambda(\e t)\sin(\theta +\alpha(\e t)) \nonumber \\
&\equiv & H_0(J) + \e^{3/2} H_1(J,\theta,\e t) , \nonumber
\eea
and the following perturbation expansion for the generating function of the canonical transformation,
\be
S(\bar J,\theta,t) = \bar J\theta + \e^{3/2} S_1 (\bar J,\theta,\e t) + \e^{5/2} S_2 (\bar J,\theta,\e t),
\ee
which allows to obtain the old action variable $J$ and the new angle variable $\bar\theta$,
\bea
J &=& \frac{\p S}{\p\theta} = \bar J + \e^{3/2} \frac{\p S_1}{\p\theta} + \e^{5/2}\frac{\p S_2}{\p\theta} \\
\bar\theta &=& \frac{\p S}{\p\bar J} = \theta + \e^{3/2} \frac{\p S_1}{\p\bar J} + \e^{5/2}\frac{\p S_2}{\p\bar J} .
\eea
After the canonical transformation, the new Hamiltonian reads
\be
\bar H (\bar J,\bar\theta,t) = H_0(J(\bar J,\bar\theta)) + \e^{3/2} H_1(J(\bar J,\bar\theta),\theta(\bar J,\bar\theta),\e t)
+ \frac{\p S}{\p t} .
\ee

Limiting to order $\e^{5/2}$ we find
\bea
H_0(J) &=& H_0\left(\bar J + \e^{3/2} \frac{\p S_1}{\p\theta} + \e^{5/2}\frac{\p S_2}{\p\theta}\right) \\
&=& H_0(\bar J) + \omega(\bar J) \left( \e^{3/2} \frac{\p S_1}{\p\theta} + \e^{5/2}\frac{\p S_2}{\p\theta}\right)
\eea
where $\omega=\frac{\p H_0}{\p J}$ is the frequency of the fast rotation.

Furthermore,
\be
H_1(J,\theta,\e t) = H_1(\bar J,\bar\theta,\e t)
\ee
and
\be
\frac{\p S}{\p t} = \e^{3/2} \frac{\p S_1}{\p t} = \e^{5/2} \frac{\p S_1}{\p (\e t)} .
\ee

Putting different terms together we obtain
\be
\bar H = H_0(\bar J) +\e^{3/2} \left[ \omega(\bar J) \frac{\p S_1}{\p \bar\theta} + H_1(\bar J,\bar\theta,\e t)\right]
+ \e^{5/2} \left[ \omega(\bar J) \frac{\p S_2}{\p\bar\theta} + \frac{\p S_1}{\p (\e t)} \right] .
\ee

Now all terms must be independent of $\bar\theta$. Since the average value of $H_1$ over the angle vanishes,
\be
\langle H_1\rangle_\theta = \frac{1}{2\pi} \int_0^{2\pi} d\theta H_1(J,\theta,\e t) = 0 ,
\ee
we must require that
\be
\omega(\bar J) \frac{\p S_1}{\p \bar\theta} + H_1(\bar J,\bar\theta,\e t) = 0,
\ee
so that
\be
\frac{\p S_1}{\p \bar\theta} = -\frac{1}{\omega(\bar J)} H_1(\bar J,\bar\theta,\e t) .
\ee

Once we have derived $S_1$ from previous relation we can impose that
\be
\omega(\bar J) \frac{\p S_2}{\p\bar\theta} + \frac{\p S_1}{\p (\e t)} = 0 .
\ee

We can easily obtain that
\be
S_1 = \frac{1}{\omega(\bar J)} \sqrt{2\bar J} \lambda\cos(\bar\theta +\alpha)
\ee
and
\be
\frac{\p S_2}{\p\bar\theta} = - \frac{1}{\omega^2(\bar J)} \sqrt{2\bar J} 
\frac{\p}{\p (\epsilon t)} \left[ \lambda\cos(\bar\theta +\alpha)\right] .
\ee

In conclusion,
\bea
\bar J &=& J -\e^{3/2} \frac{\p S_1}{\p\bar\theta} - \e^{5/2} \frac{\p S_2}{\p (\e t)} = +
J + \e^{3/2} \frac{1}{\omega(J)} H_1(J,\theta,\e t) \nonumber \\
&&+ \e^{5/2} \frac{1}{\omega^2(J)} 
\sqrt{2 J} \frac{\p}{\p (\e t)} \left[ \lambda\cos(\theta +\alpha)\right] .
\eea

Since $\omega(J) = 2J$, $J=b/b_0$, $\e =1/b_0$, and restoring the old time, we finally obtain
the adiabatic invariant
\be
b_0 \bar J =
b + \frac{1}{\sqrt{2b}} \lambda\sin(\theta+\alpha) + \frac{1}{(2b)^{3/2}} \frac{\p}{\p t}
\left[ \lambda(t) \cos (\theta+\alpha(t))\right] ,
\ee
given also in Eq.~(\ref{eq.cAI}).

\section{Energy-flux approach}
\label{app.apt}

In this appendix we describe the recursive procedure to determine the expressions of $Q_k$ based
on the energy-flux approach outlined in Section~\ref{sec.sufficient}. 
Instead of using action-angle variables as for the canonical approach (see~\ref{app.cpt}), here we
adopt the complex notation as in Eq.~(\ref{eq.UC}). Accordingly, the breather state is represented by the complex
variable $z_0(t)$ while the external signal is $w(t)=z_{-1}(t)+z_{1}(t)$. The mapping between the two sets of variables is 
\begin{eqnarray}
z_0 &=&\sqrt{b}e^{i\theta} \nonumber \\
w &=& \frac{\lambda}{\sqrt{2}} \sin\alpha + i \frac{\lambda}{\sqrt{2}} \cos\alpha ,
\end{eqnarray} 
while the evolution equation for the breather site is
\be
\label{eq:z0}
\dot z_0 = 2i |z_0|^2 z_0 + i w\,.
\ee
Given the breather  energy
\be
H_b = |z_0|^4 + z_0 w^* + z_0^* w \; ,
\ee
the energy flux is
\begin{equation}
\dot H_b = z_0 \dot w^* + z_0^*\dot w .
\label{eq.fl1}
\end{equation}
If $w$ is constant, then $H_b$ is conserved; otherwise $H_b$ can be considered as
the lowest-order approximation of the AI. In fact,
Eq.~(\ref{eq.fl1}) can be considered the first step of a recursive procedure based on the dynamical
equation
\begin{equation}
\dot Q_k = X_{k}  \; ,
\label{eq.fl2}
\end{equation}
that allows to interpret $Q_k$ as a suitable approximation of the AI ($X_k$ denotes yet unaccounted fluctuations).
The subindex $k$ in $Q_k$ ($X_k$) means that its smallest (largest) term is of order $\e^k \equiv |z_0|^{-2k}$.
By setting $k=-1/2$ in Eq.~(\ref{eq.fl2}), we recover Eq.~(\ref{eq.fl1}) (i.e., $Q_{-1/2} = H_b$, see the main text).

The recursive procedure consists in decomposing $X_k$ as the sum of two terms,
\begin{equation}
X_k = \dot q_{k+1} +  X_{k'} \qquad  k'>k
\end{equation}
in such a way that the first term in the r.h.s. captures the leading order-$k$ behavior of 
$X_k$.\footnote{Although $q_{k+1}$ is of order $k+1$, its time derivative is
of order $k$, because of the multiplicative factor due to the fast rotation.}
In practice, $q_{k+1}$ is built by integrating $X_k$ only with respect to the fast variable $z_0$.

Once $q_{k+1}$ has been identified, one can rewrite Eq.~(\ref{eq.fl2}) as
\begin{equation}
\dot Q_{k'} \equiv \dot Q_k - \dot q_{k+1} =  X_{k'}
\end{equation}
with the same structure as Eq.~(\ref{eq.fl2}) but involving a smaller residue $X_{k'}$, since $k'>k$.

By progressively increasing $k$, one can obtain increasingly accurate estimates of the AI as
\begin{equation}
Q_k = H_b - \sum_{m\le k} q_m \; .
\end{equation}
Before illustrating the various steps, it is necessary to
distinguish different harmonics of the fast breather frequency.
This is done by introducing the notation $q_k^{(l)}$, where 
the superscript $l$ identifies the oscillating (fast) frequency in units of the breather
frequency $\omega=2|z_0|^2$. 
Given that this frequency arises from the presence of the fast term  $z_0(t)=|z_0|e^{(i\omega t+\phi_0)}$ 
and its complex conjugate,
it turns out that for a given index $l$, $q_k^{(l)}$ takes the form $q_k^{(l)}=f(|z_0|)z_0^{|l|} +c.c.$, where $f$ is a generic real function.
The derivation of the first four $q_k$ is reported in the next subsections.

\subsection{The $q_{1/2}$ term}

The first step is made by decomposing $X_{-1/2}$. In this case it is sufficient to consider a single harmonic,
\be
q_{1/2}^{(1)} = -i\frac{\dot w^*}{2z_0^*} + c.c. \; .
\ee
Its time derivative can be expressed as (by invoking
the complex conjugate of Eq.~(\ref{eq.fl1}))
\be
\dot q_{1/2}^{(1)}= \left (z_0\dot w^*+ \frac{w^*\dot w^*}{2(z_0^*)^2} +c.c.\right)
- \left( \frac{i\ddot w^*}{2z_0^*} + c.c. \right)  \; .
\ee
Since $(z_0\dot w^* + c.c.)$ coincides with $\dot H_b$, we can rewrite this equation as
\be
\dot Q_{1/2} \equiv \dot H_b - \dot q_{1/2}^{(1)} = X_{1/2}^{(1)} + X_1^{(2)}
\ee
where
\be
X_{1/2}^{(1)} = \frac{i\ddot w^*}{2z_0^*} + c.c.
\ee
and
\be
X_1^{(2)} = -\frac{w^*\dot w^*}{2(z_0^*)^2} +c.c.
\ee

\subsection{The $q_{3/2}$ term}

For $k=3/2$ one harmonic again suffices, although higher harmonics appear in the expression of the remainder.
Having in mind $X_1^{(1/2)}$, we introduce
\be
q_{3/2}^{(1)} =
-\frac{\ddot w^*}{4z_0z_0^{*2}} + c.c.
\ee
Its time derivative can be expressed as 
\be
\dot q_{3/2}^{(1)} =
\frac{i\ddot w^*}{2z_0^{*}} -
\frac{iw\ddot w^*}{4z_0^2z_0^{*2}} +
\frac{iw^*\ddot w^*}{2z_0z_0^{*3}} +
\frac{\dddot w^*}{4z_0z_0^{*2}}  + c.c.
\ee
Since the first term (plus its c.c.) in the r.h.s. is equal to $X_{1/2}^{(1)}$, we can rewrite the equation as
\be
X_{1/2}^{(1)} = \dot q_{3/2}^{(1)} + X_2^{(0)} + X_2^{(2)} + X_{3/2}^{(1)}
\ee
where
\be
X_2^{(0)} = \frac{iw\ddot w^*}{4z_0^2z_0^{*2}} + c.c.
\ee
\be
X_2^{(2)} = -\frac{iw^*\ddot w^*}{2z_0z_0^{*3}} + c.c.
\ee
\be
X_{3/2}^{(1)} = -\frac{\dddot w^*}{4z_0z_0^{*2}}  + c.c.
\ee
By finally replacing the previous orders, 
\be
\dot Q_{3/2} =  X_1^{(2)} + X_{3/2}^{(1)} + X_2^{(0)} + X_2^{(2)}
\ee

\subsection{The $q_{2}$ term}

For $k=2$ two harmonics must be accounted for. 
With reference to $X_{1}^{(2)}$ we introduce
\be
q_2^{(2)} = i\frac{w^*\dot w^*}{8z_0z_0^{*3}} +c.c. 
\ee
Its derivative is
\be
\dot q_2^{(2)} =
-\frac{w^*\dot w^*}{2z_0^{*2}} +
i\frac{(\dot w^*)^2 + w^*\ddot{w^*}}{8z_0z_0^{*3}} +c.c. 
\ee
The first term in the r.h.s. coincides with $X_1^{(2)}$ and can thereby be rewritten as
\be
X_1^{(2)} = \dot q_2^{(2)} + Y_2^{(2)}
\ee
where
\be
Y_2^{(2)} = - i\frac{(\dot w^*)^2+ w^*\ddot{w^*}}{8z_0z_0^{*3}} +c.c.
\ee
Hence, the flux equation becomes,
\be
\dot H_b - \dot q_{1/2}^{(1)} - \dot q_{3/2}^{(1)} - \dot q_2^{(2)} = X_{3/2}^{(1)} + X_2^{(0)} + X_2^{(2)} + Y_2^{(2)}
\ee

Corrections of the same order arise from $X_2^{(0)}$. In this case, it is convenient to introduce
\be
q_2^{(0)} =
i \frac{w\dot w^*}{4z_0^2z_0^{*2}} + c.c.
\ee
%X_2^{(2)}
Its derivative yields
\be
\dot q_2^{(0)} =
\frac{w^2\dot w^*}{2z_0^3z_0^{*2}} 
-\frac{ww^*\dot w^*}{2z_0^2z_0^{*3}}  
+i \frac{w\ddot w^* -\dot w \dot w^*}{4z_0^2z_0^{*2}} + c.c.
\ee
%X_2^{(2)}
The second addendum in the third term does not contribute since it is purely imaginary 
(including the imaginary unit in front of the fraction).
Hence the third term coincides with $X_2^{(0)}$ and we can write
\be
X_2^{(0)} = \dot q_2^{(0)} + X_{5/2}^{(1)} + Y_{5/2}^{(1)} 
\ee
where
%X_2^{(2)}
\be
X_{5/2}^{(1)} = -\frac{w^2\dot w^*}{2z_0^3z_0^{*2}}  + c.c
\ee
\be
Y_{5/2}^{(1)} = +\frac{ww^*\dot w^*}{2z_0^2z_0^{*3}}  +c.c.
\ee
The flux equation is finally
\begin{equation}
\dot Q_{3/2} = X_{3/2}^{(1)} + X_2^{(2)} + Y_2^{(2)} + X_{5/2}^{(1)} + Y_{5/2}^{(1)} 
\end{equation}

\subsection{The $q_{5/2}$ term}

We now consider $X_{3/2}^{(1)}$ and introduce
\be
q_{5/2}^{(1)} = i\frac{\dddot w^*}{8z_0^2z_0^{*3}}  + c.c.
\ee
Its derivative is
\be
\dot q_{5/2}^{(1)} = -\frac{\dddot w^*}{4z_0z_0^{*2}}
+\frac{w \dddot w^*}{8z_0^3z_0^{*3}} 
-\frac{3w^*\dddot w^*}{8z_0^2z_0^{*4}}  
+i\frac{\ddddot w^*}{8z_0^2z_0^{*3}}  + c.c.
\ee
The first term is $X_{3/2}^{(1)}$ and can thereby be rewritten as
\be
X_{3/2}^{(1)} = \dot q_{5/2}^{(1)} + X_3^{(0)} + X_3^{(2)} + Z_{5/2}^{(1)}
\ee
where
\be
X_3^{(0)} = -\frac{w \dddot w^*}{8z_0^3z_0^{*3}} c.c
\ee
\be
X_3^{(2)} = \frac{3w^*\dddot w^*}{8z_0^2z_0^{*4}}  + c.c.
\ee
\be
Z_{5/2}^{(1)} = - i\frac{\ddddot w^*}{8z_0^2z_0^{*3}}  + c.c.
\ee
Finally, the flux equation at this order of approximation can be written as
\begin{equation}
\dot Q_{5/2} = X_2^{(2)} + Y_2^{(2)} + X_{5/2}^{(1)} + Y_{5/2}^{(1)} + X_3^{(0)} + X_3^{(2)} + Z_{5/2}^{(1)}
\end{equation}

\section*{References}

\bibliographystyle{iopart-num}

\bibliography{biblio.bib}

\end{document}